\documentclass[aps,prd,showpacs,nobibnotes,twocolumn,
nofootinbib,nobalancelastpage,superscriptaddress,preprintnumbers]{revtex4}
\usepackage{graphicx}
\usepackage{amsmath,color}
\usepackage{latexsym}
\usepackage{dcolumn}
\usepackage{bm}
\input{epsf}
\newcommand{\be}{\begin{equation}}
\newcommand{\ee}{\end{equation}}
\newcommand{\ba}{\begin{eqnarray}}
\newcommand{\ea}{\end{eqnarray}}
\newcommand{\bi}{\begin{itemize}}
\newcommand{\ei}{\end{itemize}}

\newcommand{\bfi}{\begin{figure}
\epsfxsize=9cm
\epsffile}
\newcommand{\efi}{\end{figure}}

\newcommand{\WMAP}{\emph{WMAP}}

\begin{document}
\title{IceCube and HAWC constraints on very-high-energy \\ emission from the \emph{Fermi} bubbles}
\author{Ke Fang}
\affiliation{University of Maryland, Department of Astronomy, College Park, MD, 20742}
\affiliation{Joint Space-Science Institute, College Park, MD, 20742}
\author{Meng Su}
\affiliation{Department of Physics and Laboratory for Space Research, the University of Hong Kong, PokFuLam, Hong Kong SAR, China}

\author{Tim Linden}
\affiliation{Ohio State University, Center for Cosmology and AstroParticle Physics (CCAPP), Columbus, OH 43210}
\author{Kohta Murase}
\affiliation{Center for Particle and Gravitational Astrophysics, The Pennsylvania State University, University Park, PA 16802, USA}
\affiliation{Department of Astronomy \& Astrophysics, The Pennsylvania State University, University Park, PA 16802, USA
}
\affiliation{Department of Physics, The Pennsylvania State University, University Park, PA 16802, USA}
\affiliation{Yukawa Institute for Theoretical Physics, Kyoto University, Kyoto 606-8502, Japan} 
\begin{abstract}
The nature of the $\gamma$-ray emission from the \emph{Fermi} bubbles is unknown. Both hadronic and leptonic models have been formulated to explain the peculiar $\gamma$-ray signal observed by the Fermi-LAT between 0.1-500~GeV. If this emission continues above $\sim$30~TeV, hadronic models of the \emph{Fermi} bubbles would provide a significant contribution to the high-energy neutrino flux detected by the IceCube observatory. Even in models where leptonic $\gamma$-rays produce the \emph{Fermi} bubbles flux at GeV energies, a hadronic component may be observable at very high energies. The combination of IceCube and HAWC measurements have the ability to distinguish these scenarios through a comparison of the neutrino and $\gamma$-ray fluxes at a similar energy scale. We examine the most recent four-year dataset produced by the IceCube collaboration and find no evidence for neutrino emission originating from the \emph{Fermi} bubbles. In particular, we find that previously suggested excesses are consistent with the diffuse astrophysical background with a p-value of 0.22 (0.05 in an extreme scenario that all the IceCube events that overlap with the bubbles come from them). Moreover, we show that existing and upcoming HAWC observations provide independent constraints on any neutrino emission from the \emph{Fermi} bubbles, due to the close correlation between the $\gamma$-ray and neutrino fluxes in hadronic interactions. The combination of these results disfavors a significant contribution from the \emph{Fermi} bubbles to the IceCube neutrino flux.
\end{abstract}
\maketitle

\section{Introduction}
\label{introduction}
Observations by the Fermi Large Area Telescope (Fermi-LAT) have discovered two extended ``bubbles" filling the regions above and below the Galactic center~\cite{FermiBubble}. Assuming that these bubbles originate in the Galactic center, their angular size translates to a physical extent exceeding $\sim$10~kpc. Intriguingly, the $\gamma$-ray spectrum of the \emph{Fermi} bubbles does not show significant variation up to $\sim$50$^\circ$ away from the Galactic plane. This suggests that the particles producing the \emph{Fermi} bubbles emission do not cool significantly during cosmic-ray transport. Additionally, the morphology of the bubbles shows sharp edges suggestive of a transient origin. 

The \emph{Fermi} bubbles are connected to previous Wilkinson Microwave Anisotropy Probe (WMAP) observations of a bright excess in the inner galaxy, known as the \WMAP\ ``haze"~\cite{Finkbeiner:2003im, Dobler:2008ww}. Subsequent observations of the \WMAP\ haze have verified this correlation, finding that the sharp edges of the \emph{Fermi} bubbles are also seen in the haze emission~\cite{Dobler:2011rd}. The spectrum of the \WMAP\ haze is softer than free-free emission, but is harder than synchrotron emission from astrophysical electrons produced through standard diffusive processes. Models typically produce the \WMAP\ haze via synchrotron emission from an unknown, hard-spectrum cosmic-ray (CR) electron population~\citep{FermiBubble}. The morphological similarity of the \WMAP\ haze and \emph{Fermi} bubbles suggests a similar leptonic origin for the bubbles. This CR lepton population could be formed through an episodic event that produced an intensive energy injection near the Galactic center, such as a past accretion event onto Sgr A*, or a nuclear starburst~\citep{2012ApJ...756..181G, 2012ApJ...756..182G, 2012ApJ...761..185Y, 0004-637X-778-1-58}. 

However, TeV leptons interact with both the magnetic field and interstellar radiation field (ISRF) of the Milky Way, losing energy to synchrotron radiation and inverse-Compton scattering on timescales of only $\sim$0.2~Myr. Unless the TeV electrons were injected very recently, leptonic models would generically include a significant softening in the \emph{Fermi} bubbles $\gamma$-ray spectrum at high energies, which is not confirmed due to limited statistics at those energies. The process that produces a consistent $\gamma$-ray spectrum throughout the \emph{Fermi} bubbles is unknown, and may hold the key to understanding the origin of the bubbles. Two classes of models have been proposed.

In the first class of models, the $\gamma$-ray signal is created by primary electrons accelerated near the Galactic center. In this scenario, the hard spectrum of the \emph{Fermi} bubbles is produced by modifying \emph{cosmic-ray transport.} If CRs are transported via diffusion, a diffusion coefficient of $\sim$2.5$\times$10$^{31}$~cm$^2$s$^{-1}$ at 1~TeV would be required. This value exceeds the average Milky Way diffusion coefficient by nearly two orders of magnitude. A strong convective wind in the Galactic center could be employed to more effectively move electrons out of the Galactic plane. However for 1~TeV electrons, this corresponds to an unphysical wind speed of $\sim$50,000~km/s~\citep{FermiBubble}. One alternative possibility involves slower CR electron transport coupled with significant re-acceleration forces (e.g. due to Alfv{\'e}n waves) which offset energy losses and keep the electron spectrum in steady state~\citep{0004-637X-790-1-23, PhysRevLett.107.091101,2015ApJ...814...93S}. It has also been proposed that CRs could be injected by shocks near the bubble edge and diffuse away from it at high energies \citep{2016arXiv161104190K}.

In the second class of models, the primary CR population is energetically dominated by hadrons. The cooling of hadronic CRs is significantly slower than leptons, with a characteristic cooling time of \mbox{$t_{\rm pp}= \left(n_{\rm H}\,\sigma_{\rm pp}\,c\right)^{-1} \sim 1.8 \times10^3 \,({\rm n_H}/10^{-2}\,{\rm cm}^{-3})^{-1}~{\rm Myr}$}, where ${n_H}$ is the number density of the cold gas within the bubble, and $\sigma_{\rm pp}\sim 6\times10^{-26}\,\rm cm^2$ is the inelastic part of the pp cross section at PeV energies \citep{PhysRevD.74.034018}. This allows hadronic CRs to easily propagate to the outer edges of the bubbles without significant energy losses. The hadronic interactions of these CRs produce an energetic CR lepton population {\it in situ}. This solves the CR transport problem at the cost of invoking an intermediate stage. 

There are two important implications of models utilizing hadronic cosmic-ray transport to explain the \emph{Fermi} bubbles. First, the flux of $\gamma$-rays and e$^+$e$^-$ pairs produced in hadronic interactions are similar. This indicates that the majority of the observed $\gamma$-ray emission from the \emph{Fermi} bubbles is, in fact, produced via the decay of neutral pions formed in the initial hadronic interaction. The contribution of inverse-Compton scattering from the lepton population is subdominant. Additionally, models find that the synchrotron emission in a hadronic scenario is typically 3-4 times lower than WMAP and Planck measurements \citep{FB2014}. Hadronic models of the \emph{Fermi} bubbles thus require a sub-dominant component of primary electrons, either accelerated {\it in situ} or transported by Galactic winds, in order to explain the \WMAP\ emission~\citep{Crocker:2011, 2014ApJ...789...67F}.

Second, while energy losses constrain the maximum energy of leptons in the \emph{Fermi} bubbles to be $\sim 1-10$ TeV-scale, protons can be accelerated to much higher energies. Although the CR acceleration site is uncertain, hadronic models often extend the \emph{Fermi} bubbles spectrum to PeV energies. Supernova remnants are believed to accelerate CRs up to PeV energies, and CRs may be injected by the past starburst activities \citep{Crocker:2011}. Alternatively, SgrA$^*$ may be an efficient PeV accelerator, and is expected to have a maximum energy output sufficient to produce the bubbles during a transient event~\citep{2016arXiv160400003F}. Another possibility is that protons may be accelerated at termination shocks around the edge of the \emph{Fermi} bubbles, the energy of these accelerated protons may reach PeV energies in $\mu$G magnetic fields, although the maximum energy can also be much smaller~\citep{2014MNRAS.444L..39L, 2014ApJ...789...67F}. 


While the limited field-of-view of atmospheric Cherenkov Telescopes makes them incapable of cosntraining the $\sim$TeV emission from the \emph{Fermi} bubbles, the high sensitivity and wide field-of-view offered by the High Altitude Water Cherenkov (HAWC) telescope is capable of testing the extension of the \emph{Fermi} bubbles to the TeV regime. Early observations by the HAWC collaboration have found no evidence for TeV $\gamma$-ray emission originating from the \emph{Fermi} bubbles~\citep{HAWC_UL}.

Hybrid models are also possible. In particular, models with a significant primary electron component typically accelerate a population of very-high-energy (VHE) protons as well. In any hadronic or hybrid model a correlation is expected between the $\gamma$-ray and neutrino fluxes. In particular, any hadronic $\gamma$-rays must be accompanied by high-energy neutrinos emitted via the decays of charged pions produced in the hadronic interaction. On the other hand, leptonic models produce no associated neutrino emission. Thus, a discovery or non-detection of high-energy neutrinos offers the potential to differentiate hadronic and leptonic models of the \emph{Fermi} bubbles~\citep{2012PhRvL.108v1102L}. Meanwhile, the origin of the high-energy neutrinos observed by the IceCube Observatory remains a mystery \citep{Halzen:2016gng}. Because the \emph{Fermi} bubbles are bright at GeV energies, they have been suggested as potential contributors to the TeV neutrino sky~\citep{Crocker:2011}. An excess around the Galactic center was tentatively indicated in the two-year high-energy starting event (HESE) data \citep{1242856}, and a possible contribution of Fermi bubbles to the diffuse neutrino flux has been discussed \citep{2014PhRvD..90b3010A,2014PhRvD..90b3016L,  2014PhRvD..89j3003T,2015EPJC...75..116A,  2015PhRvD..92b1301L,  2016PhRvD..93a3009A}. It has been suggested that both neutrino observations with IceCube as well as multi-TeV $\gamma$-ray observations with HAWC and other air-shower arrays are crucial to test hadronic models of the \emph{Fermi} bubbles \citep{2014PhRvD..90b3010A, 2015PhRvD..92b1301L}.

In this paper, we re-analyze the flux of neutrinos from the {\it Fermi} bubbles, utilizing the most recent four-year IceCube constraints as well as HAWC observations of the TeV $\gamma$-ray emission from the bubbles. By considering both atmospheric and astrophysical neutrino backgrounds in the \emph{Fermi} bubbles region, we show that there is no statistically significant excess of high-energy neutrinos correlated with the \emph{Fermi} bubbles. Additionally, we show that HAWC results independently rule out models where a significant IceCube neutrino flux is produced in a  purely hadronic model of the Fermi bubbles. While hybrid models are still possible, we show that future HAWC observations could exclude  at least half of the overlapping neutrino events as having a \emph{Fermi} bubbles origin.

The paper is organized as follows. In Section~\ref{sec:flux} we calculate the neutrino flux within the \emph{Fermi} bubbles region, and use models for the atmospheric and astrophysical backgrounds to constrain the neutrino excess associated with the {\it Fermi} bubbles. In Section~\ref{sec:spec} we consider constraints from HAWC null-observations of the \emph{Fermi} bubbles in both hadronic and hybrid leptonic/hadronic models. Finally, in Section~\ref{sec:discussion} we discuss the prospects for constraining the neutrino emission associated with the \emph{Fermi} bubbles with current and future experiments.

\section{IceCube Observations of the \emph{Fermi} bubbles}
\label{sec:flux}

\begin{figure}[t!]
\begin{center}
\includegraphics[width=.49\textwidth]{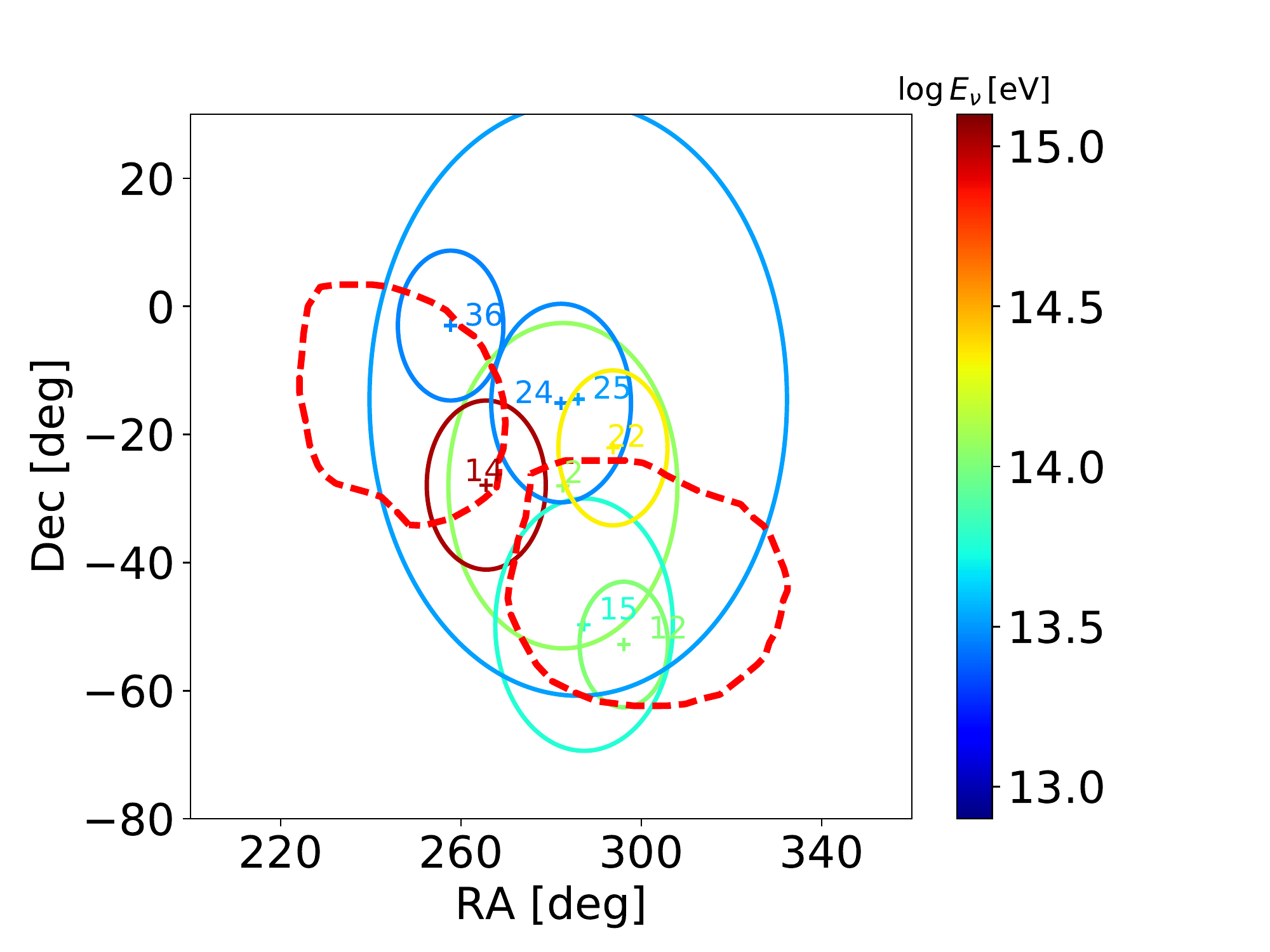}
\end{center}
\caption{The spatial distribution of neutrino events in the IceCube four-year HESE data \citep{IC3yr, IC4yr} that overlap with the \emph{Fermi} bubbles \citep{FB2014}. The results are shown in equatorial coordinates. Neutrino events are labeled with their event ID. The contour surrounding each event corresponds to its angular resolution. Eight events from the four-year data partially overlap with the \emph{Fermi} bubbles. \mbox{Events 2, 12, 14, 15, and 36} have best-fit arrival directions that lie inside the bubbles, while \mbox{events 22, 24, and 25} are centered outside the bubbles. The event distribution shown here is identical to the results of the three-year data~\cite{2015PhRvD..92b1301L}, since no new events from the fourth-year of data overlap with the bubbles.}
\label{fig:bubble}
\end{figure}

In four years of data (1347 days), observations by the IceCube Observatory have detected 54 neutrinos with verticies inside the IceCube detector and contained energies exceeding $\sim$30~TeV. These events are known as the high-energy starting events (HESE)~\citep{1242856, IC3yr, IC4yr}.  Of these events, 8 spatially overlap with the \emph{Fermi} bubbles, as shown in Figure~\ref{fig:bubble}. For each event we show an ellipse that corresponds to the angular uncertainty in the event reconstruction, and an event color that corresponds to the reconstructed energy of the neutrino. All of these events are shower events, which have a good energy resolution ($\sim$15\%), but suffer from large uncertainties in the reconstruction of their arrival direction ($\sim$10$^\circ$). This is comparable to the size of the \emph{Fermi} bubbles, and implies that all events near the location of the \emph{Fermi} bubbles must be studied in detail. In particular, we find that five events (2, 12, 14, 15, and 36) have a reconstructed direction that is centered within the bubbles. An additional three events (22, 24, 25) are in close proximity, but have reconstructed directions that are centered outside the bubbles.  Event 14, which is located close to the Galactic center, is the highest-energy event in the region and is one of only three events in the full HESE dataset with reconstructed energies exceeding 1~PeV.

The poor angular resolution of these HESE events makes it difficult to evaluate the spatial correlation between each event and the \emph{Fermi} bubbles. In what follows, we utilize the best-fit directions of each event, and consider only candidate events that are centered within the \emph{Fermi} bubbles region. We note that weighting each event by the fraction of the point-spread function that lies within the \emph{Fermi} bubbles provides similar results. We take the solid angle of the \emph{Fermi} bubbles to be \mbox{$\Omega_{\rm FB} \approx 0.85$~sr~\citep{FB2014}}. The total number of events in the bubbles is $N_{\rm FB}^{\rm tot}=5$.

\begin{figure}[t]
\begin{center}
\includegraphics[width=.5\textwidth]{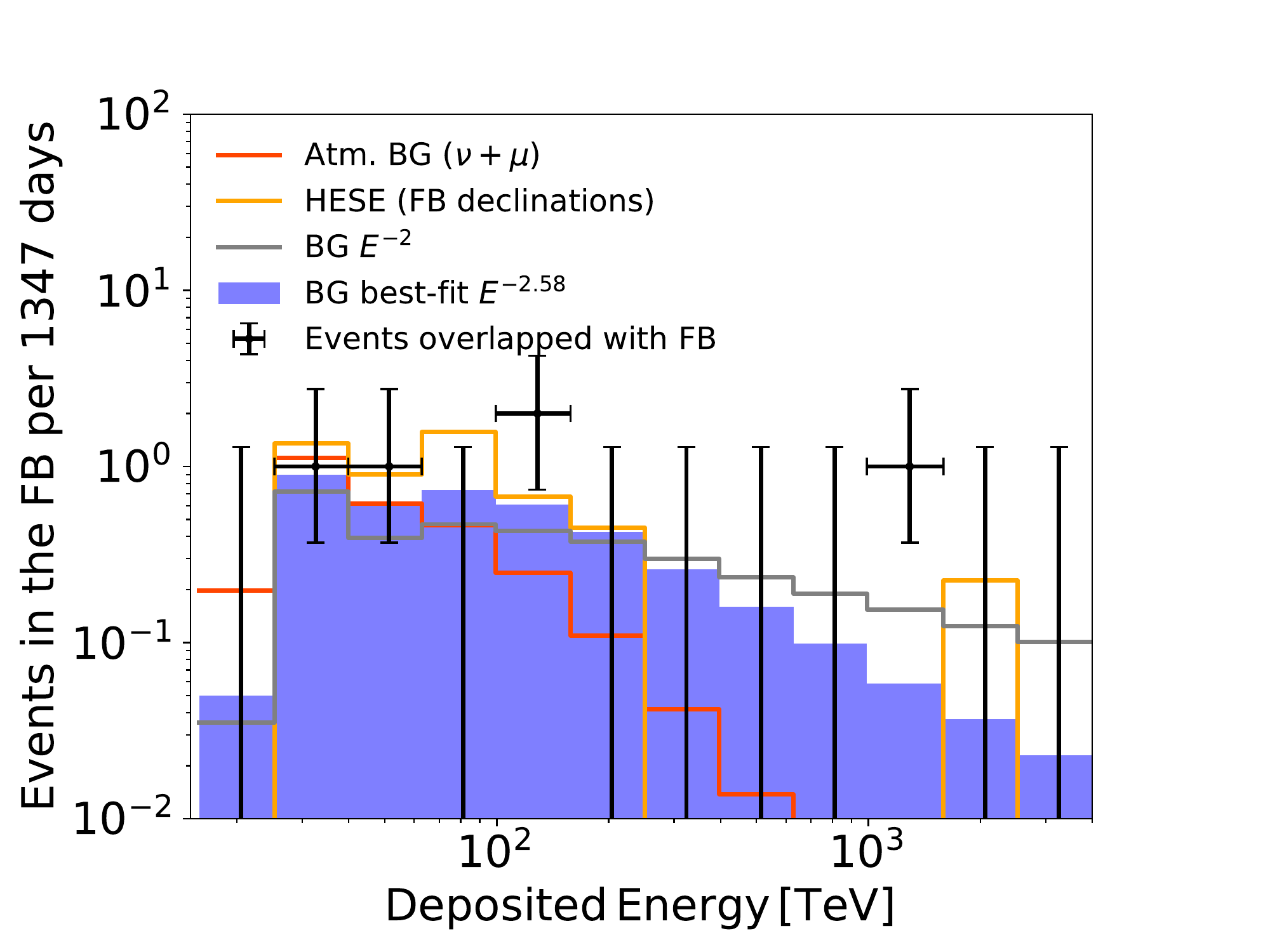}
\end{center}
\caption{The number of neutrinos in the \emph{Fermi} bubbles as a function of their deposited energy. Results are based on HESE events observed over four years of IceCube data~\citep{IC4yr}. The dataset is compared to: (1) the number of expected atmospheric events including neutrinos and muons (solid red), (2) the average number of events in the declination range spanned by the \emph{Fermi} bubbles \mbox{($-60^\circ<\delta < 0^\circ$; solid orange)}, and (3) the predicted number of atmospheric and astrophysical neutrinos between 60~TeV and 3~PeV based on the isotropic data. We utilize two models for the spectrum of astrophysical neutrinos, including an E$^{-2}$ spectrum (grey), and an E$^{-2.58}$ spectrum fit to the IceCube HESE (blue shaded region). A total event number of 5.2 is expected based on event distribution in the sideband regions, which is comparable to the observed number of 5. The flux and distribution of neutrinos in the \emph{Fermi} bubbles is also consistent with the best-fit model of the isotropic sky, with an excess that has a p-value of 0.22.}
\label{fig:Nevent}
\end{figure}

In Figure~\ref{fig:Nevent} we show a distribution of the deposited energy for events centered within the \emph{Fermi} bubbles region. The error bars are calculated using Feldman-Cousins confidence intervals~\cite{1998PhRvD..57.3873F}. We compare these events against the expected number of events within the bubbles from atmospheric backgrounds (red), including atmospheric neutrinos from $\pi$/K decays, charmed meson decays, and background muons~\citep{IC4yr}, as well as the expected number of HESE events in the ROI from observations of sideband regions in a similar declination range~($-60^\circ<\delta<0^\circ$; orange). 

Additionally, we show the sum of the expected atmospheric background along with a best-fit astrophysical diffuse background, which is estimated by:
\begin{equation}
    N_{\rm FB}^{\rm B} = N_{\rm south}^{\rm B} \,\frac{\Omega_{\rm FB}}{2\,\pi},
\end{equation}
where $N_{\rm south}^{\rm B}$ is the expected background number in the Southern sky, scaled from the all-sky background using the average effective area of the Northern and Southern sky \citep{1242856}, 
\begin{equation}
    N_{\rm south}^{\rm B} = N_{\rm all}^{\rm B}\,\frac{A_{\rm eff, south}}{A_{\rm eff,south}+A_{\rm eff,north}}.
\end{equation}
$N_{\rm all}^{\rm B}$ is the expected all-sky background number, obtained using the best-fit spectral slope of E$^{-2.58}$ obtained from all HESE neutrino events with a deposited energy between 60~TeV and 3~PeV~\citep{IC4yr}. We also show a comparison to the background number obtained with a fixed $E^{-2}$ spectrum (indicated by the grey steps). 
Note that we have adopted the Southern sky (where the bubbles are located) as  a  reference, since the effective area of the IceCube detector   is   different for upgoing (Northern sky) and downgoing (Southern sky) events \footnote{As a cross-check, we note that using the average effective area of the Southern sky, the best-fit isotropic per-flavor neutrino flux of $0.84\times10^{-8}\,\rm GeV^{-1}\,cm^{-2}\,s^{-1}\,sr^{-1}$~\citep{IC4yr} predicts the observation of 35.5 events in the Southern sky. This is comparable to the actual value of 37.}. 

By comparing our background models with the observed neutrino counts from the \emph{Fermi} bubbles region, we find that there is no significant excess coincident with the bubbles. While the number of neutrinos observed in the \emph{Fermi} bubbles region significantly exceeds the atmospheric background, the flux within the \emph{Fermi} bubbles is consistent with the summed contribution from atmospheric and astrophysical backgrounds. In particular, the \emph{Fermi} bubbles neutrino flux is well-fit by both background models that are derived from the sidebands regions at similar declinations, as well as the observed isotropic neutrino flux.

To quantify the significance of any excess from the \emph{Fermi} bubbles region, we compare the number of events observed in each energy bin to the number of neutrinos expected based on the combination of atmospheric and astrophysical neutrino backgrounds. We define a test statistic (TS), calculated as the sum of the logarithms of the probability of finding the observed number of neutrinos in each bin E$_{i}$, assuming that the number of neutrinos at each deposited energy is independent and follows a Poisson distribution:

\begin{equation}
   {\rm TS} = \sum_i\left[{\rm log}\left(p(E_i)\right)\right]
\end{equation}

To determine the distribution of TS values expected from Poisson fluctuations in the background model, we utilize Monte Carlo techniques to produce a large population of mock observations, based on fluctuations of the background model. We find that in 22\% of these observations, the resulting TS exceeds that computed for the observed data. Considering that most of the nearby events are shower events with large uncertainties, we also test an extreme scenario that all overlapped events are from the \emph{Fermi} bubbles. Even with this assumption, we still find that 5\% of realizations with background-only hypothesis resulted a TS higher than the observed value. We thus conclude that the neutrino flux from the \emph{Fermi} bubbles region is consistent with background fluctuations.

\begin{figure*}[ht]
\begin{center}
\includegraphics[width=.9\textwidth]{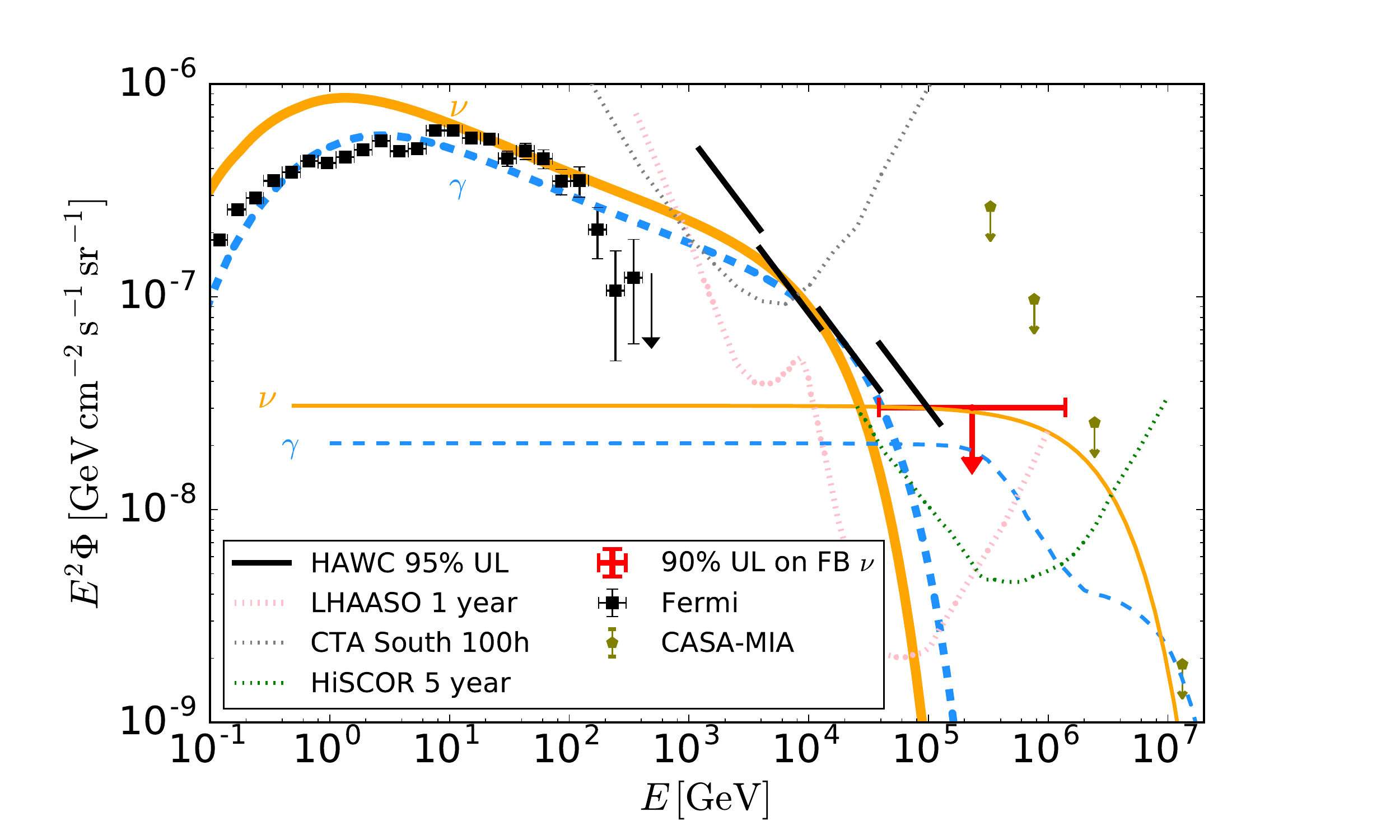}
\end{center}
\caption{The modeled intensity and spectrum of the neutrino and $\gamma$-ray emission produced by hadronic interactions in the {\it Fermi} bubbles. We show the predicted $\gamma$-ray (blue dashed) and all-flavor neutrino (orange solid) spectrum for our models of hadronic \emph{Fermi} bubbles production (thick lines), as well as the hadronic fraction of our hybrid leptonic-hadronic model (thin lines). Details of the models are given in Section~\ref{sec:spec}. We note that the $\gamma$-ray spectrum in our leptonic-hadronic model receives additional contributions from the interactions of primary electrons, which are not shown here. We compare our results to $\gamma$-ray observations of the \emph{Fermi} bubbles by the Fermi-LAT at GeV energies (black squares), the 95\% confidence upper limits on the TeV $\gamma$-ray flux recorded by HAWC (black solid bars), the 90\% confidence upper limits on ultrahigh-energy gamma rays by CASA-MIA scaled to the bubbles region (olive upper limits; \cite{1997PhRvL..79.1805C, 2014PhRvD..90b3010A}), and the $90\%$ confidence upper limit on the neutrino flux at TeV---PeV energies as calculated in this work (red upper limit). We additionally show the projected sensitivity from 100~hr of CTA observations (grey dotted; \citep{Lemoine-Goumard:2015cna}), 5~yr of HiSCOR observations (green dotted; \citep{2012NIMPA.692..246T}), and 1~yr of LHASSO observations (pink dotted; \citep{doi:10.1142/S2010194512005867}) converted to the region of the \emph{Fermi} bubbles following \cite{2014PhRvD..90b3010A}, assuming that these detectors would be able to view (or have viewed) the Fermi bubbles continuously for assumed periods. In the hadronic scenario (thick lines), the maximum neutrino flux allowed by the Fermi-LAT and HAWC measurements does not produce a significant IceCube flux at high neutrino energies. However, in the hybrid leptonic-hadronic scenario (thin lines), the spectral index of the sub-dominant $\gamma$-ray component can be extremely hard, producing a bright neutrino flux detectable by IceCube. We note that the IceCube upper limit is calculated over a wide energy bin, and a significant number of neutrinos are observed at energies exceeding $\sim$100~TeV where the flux in the pure hadronic model is negligible.}
\label{fig:spec}
\end{figure*}

We find that the best-fit flux from the background model accounts for $3.97$ events in the \emph{Fermi} bubbles region. This implies that the best-fit flux of the \emph{Fermi} bubbles (while not statistically significant), accounts for $N_{\rm FB} = 1.03$ events. Utilizing Feldman-Cousins statistics, we can set a 90\% confidence upper limit on the total contribution of the \emph{Fermi} bubbles to the IceCube neutrino flux of 5.99 predicted events~\cite{1998PhRvD..57.3873F}. Of the $3.97$ events expected from our background model, we find that $1.6$ events are produced by the atmospheric background, while $2.37$ events are produced by the isotropic astrophysical background.

The upper limit for the neutrino flux from the \emph{Fermi} bubbles can then be estimated from the all-sky astrophysical flux through  \citep{2014PhRvD..90b3010A}:
\begin{eqnarray}\label{eqn:Flux}
 {J_{\rm FB}^{\rm all-flavor}}& \leq & \frac{N_{\rm FB}^{\rm UL}}{N_{\rm south}}\,\frac{2\,\pi}{\Omega_{\rm FB}}\,3\,J_{\rm IC} \\ \nonumber
 &=& 3.0\times10^{-8}\,\rm GeV\,cm^{-2}\,s^{-1}\,sr^{-1},
\end{eqnarray}
where $N_{\rm south} = 37$ is the Southern-sky HESE event number in the four-year data, and    $J_{\rm IC} =\left( 0.84\pm0.3\right) \times10^{-8}\,\rm GeV\,cm^{-2}\,s^{-1}\,sr^{-1}$ is the per-flavor astrophysical flux obtained for events between 60 TeV and 3 PeV assuming a fixed $E^{-2}$ spectrum \citep{IC4yr}.

\section{HAWC Observations of the \emph{Fermi} bubbles}\label{sec:spec}

Recently, the HAWC collaboration analyzed the $\gamma$-ray emission from the \emph{Fermi} bubbles region over 290~days of HAWC observation time. They found no statistically significant excess, and produced strong constraints on the $\gamma$-ray flux between energies of 1-100~TeV~\citep{HAWC_UL}. Utilizing these results, the HAWC collaboration placed stringent upper limits on the maximum neutrino flux in the \emph{Fermi} bubbles, within the context of purely hadronic models. Interestingly, these observations ruled out a previous analysis of the neutrino flux from the \emph{Fermi} bubbles, which did not take into account the possibility that the neutrino flux could be produced by atmospheric or astrophysical backgrounds~\citep{Lunardini:2014}.

Here we generalize the results of the HAWC collaboration, and consider the potential for HAWC data to place model-independent constraints on the contribution of the \emph{Fermi} bubbles to the IceCube neutrino flux. In particular, we consider two scenarios, which we call ``pure hadronic" and ``hybrid leptonic-hadronic". In the pure hadronic scenario, we assume that all $\gamma$-rays produced in the \emph{Fermi} bubbles at GeV energies are produced via hadronic processes. In this case, the detection of bright \emph{Fermi} bubbles emission by the Fermi-LAT, coupled with the strong upper limits on \emph{Fermi} bubbles emission by HAWC, combine to force the $\gamma$-ray spectrum (and by extension the neutrino spectrum) to be extremely soft. In the hybrid leptonic-hadronic scenario we instead assume that the bulk of the GeV $\gamma$-ray signal observed by the Fermi-LAT is produced by primary leptons, while $\gamma$-rays from hadronic processes are sub-dominant. This allows the spectrum of hadronic $\gamma$-rays and neutrinos to be relatively hard, allowing for a larger very-high-energy flux.

In each case, we fit the $\gamma$-ray spectrum and intensity to Fermi-LAT and HAWC observations, and then calculate the resulting neutrino spectrum under the assumption that neutrinos and $\gamma$-rays from hadronic interactions are correlated via the relationship \citep{2014PhRvD..90b3010A}:
 \begin{equation}
    \label{eq:neutrinospectrum}
     \left(E_\nu\,Q_{E_\nu}\right)_{\rm all-flavor} \approx \frac{3}{2}\,\left(E_\gamma\,Q_{E_\gamma}\right)|_{E_\nu = E_\gamma/2},
 \end{equation}
where $Q_E\propto E\,d{\dot N}/dE$ is the production rate of neutrinos and $\gamma$-rays. The number of neutrino events in the bubble region can then be calculated by:

\begin{equation}
    N_{\nu} = \int_{30\,\rm TeV}^{\rm 1 PeV} \left(\frac{dN}{dE\,dA\,dt\,d\Omega}\right)_\nu \,A_{\rm eff,south}\,t_{\rm live}\,\Omega_{\rm FB},
\end{equation}
where $A_{\rm eff}$ is the effective area of IceCube for contained  neutrino searches averaged over the Southern sky \citep{1242856}. This serves as a reasonable approximation for the average effective area of IceCube in the \emph{Fermi} bubbles region.

In Figure~\ref{fig:spec} we show the results of this analysis. We compare our models to four datasets: the 0.1-500~GeV measurements of the \emph{Fermi} bubbles by the Fermi-LAT collaboration (black squares~\citep{Fermi14}), the 95\% confidence upper limits from the non-detection of very-high-energy $\gamma$-rays in the northern bubble by HAWC (black solid bars), the 90\% confidence upper limits from the non-detection of ultrahigh-energy $\gamma$-rays by the Chicago Air Shower Array - Michigan Muon Array experiment (CASA-MIA, olive upper-limits \citep{1997PhRvL..79.1805C}), and the neutrino flux calculated in this paper using 4-year HESE data (red upper-limit). Because the neutrino data is extremely sparse, we combine all energy bins in our neutrino analysis into one energy bin, depicted at a central energy of  $\sim$230~TeV. We have included uncertainties in the neutrino flux stemming from the difference between the true neutrino energy and the deposited energy following \cite{2014arXiv1408.3799B, 2013PhRvD..88d3009L}. We show the predicted $\gamma$-ray (blue dashed) and neutrino (orange solid) fluxes from the pure hadronic scenario (thick lines), as well as the hadronic portion of the emission in the hybrid hadronic-leptonic scenario (thin lines). In the next three subsections, we will investigate these models in detail, using the latest HAWC and IceCube constraints.

\subsection{Pure Hadronic Models of the \emph{Fermi} bubbles}\label{sec:had}

\begin{figure}[ht]
\begin{center}
\includegraphics[width=.5\textwidth]{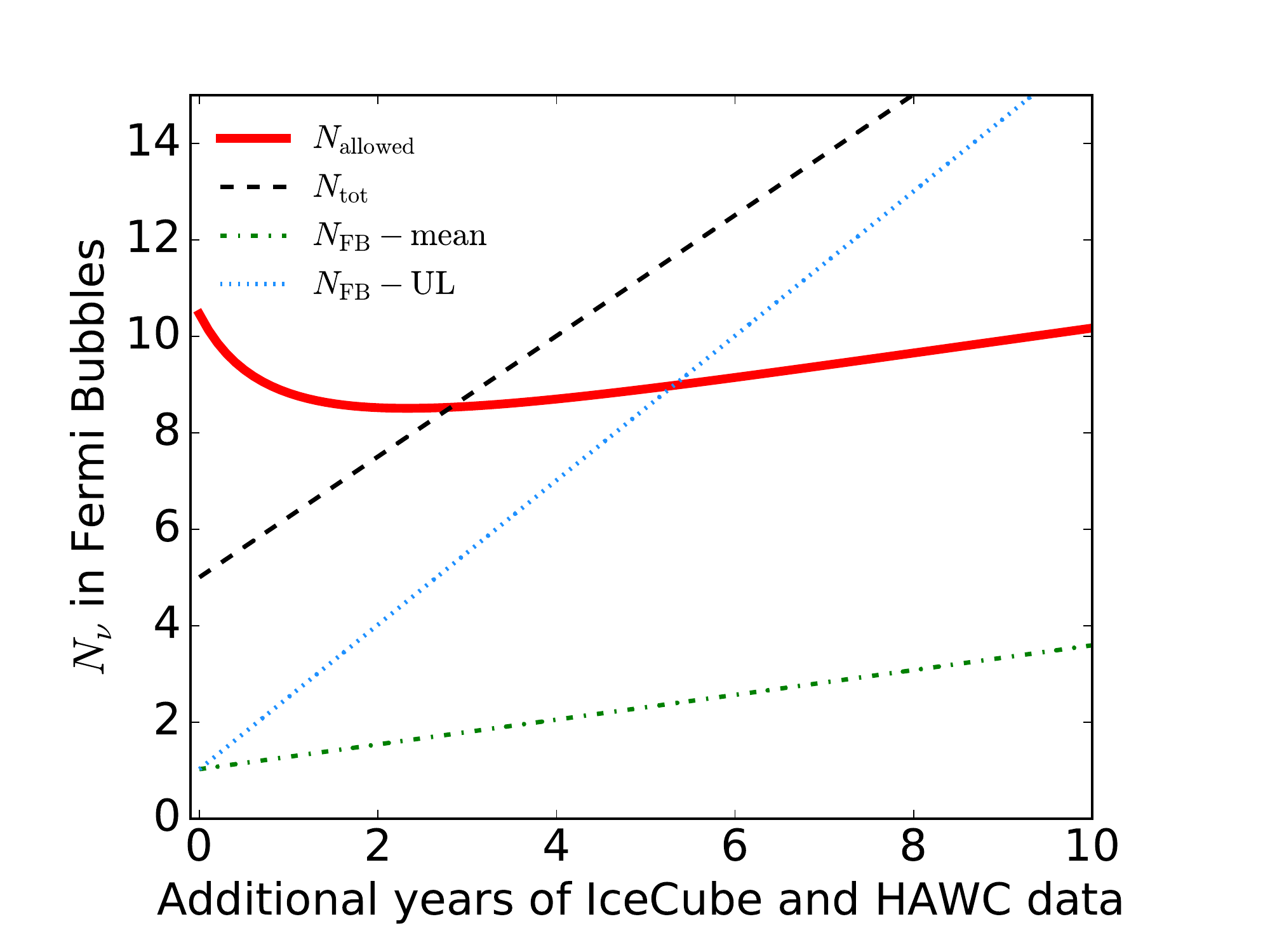}
\end{center}
\caption{Projected HAWC upper limits on the contribution of the \emph{Fermi} bubbles to the IceCube neutrino flux, as a function of the number of additional years of HAWC and IceCube data. The red solid line indicates the maximum number of neutrinos produced by the \emph{Fermi} bubbles that would remain consistent with null-observations of the bubbles by HAWC. These results utilize our hybrid leptonic-hadronic model. We show three IceCube predictions, including: (1) the total number of neutrinos in the \emph{Fermi} bubbles region, extrapolated from HESE data (black dashed), (2) the extrapolation of the best-fit neutrino flux derived in this paper, which produces $N_{\rm FB} = 1.03$ events in the HESE dataset (green dash-dotted), and (3) the extrapolation of the current IceCube 90\% confidence upper-limit on the neutrino contribution from the bubbles, which would be expected to produce $N_{\rm FB}^{\rm UL}=5.99$ events over four years of neutrino data (blue dotted). After five more years of data, HAWC could begin to constrain the neutrino flux originating from the bubbles. Observations with seven additional years of HAWC and IceCube data could conclusively show that no more than $\sim$50\% of the IceCube neutrino flux in the bubbles region is due to the \emph{Fermi} bubbles. We have assumed an E$^{-2}$ spectrum for the $\gamma$-ray and neutrino signal in the leptonic-hadronic model. A softer spectrum would produce more stringent limits. The x-axis denotes the additional year after the 4-year HESE data for IceCube \citep{IC4yr}, and after the 290 days of observations for HAWC \citep{HAWC_UL}.}
\label{fig:Nyr}
\end{figure}

In pure hadronic models of the {\it Fermi} bubbles, the $\gamma$-ray spectrum is fit through a comparison of the bright $\sim$100~GeV emission observed by the Fermi-LAT, compared with the stringent upper limits on $\sim$10~TeV $\gamma$-ray emission produced by HAWC. We employ a proton spectrum modeled as a power-law with an exponential cutoff \mbox{(i.e. $dN/dE\propto E^{-\alpha}\,\exp(-E/E_{\rm cut})$)}. We adopt the best-fit hadronic model with index $\alpha\sim 2.2$ as found by\cite{FB2014}. We then saturate the HAWC upper limit at $\sim$10~TeV. We find that this requires an exponential cutoff in the bubbles spectrum at an energy $E_{\rm cut}\lesssim  50$~TeV. 
We note that we could theoretically soften the injected $\gamma$-ray proton spectrum to allow higher values of E$_{\rm cut}$, which would consequently allow for a higher PeV neutrino flux. 
However, a softer $\gamma$-ray spectrum would provide a poor fit to the Fermi-LAT data.

Translating this $\gamma$-ray flux into a neutrino flux following Equation~\ref{eq:neutrinospectrum}, we find that IceCube should detect no more than $N_{\nu,\rm H}= 0.7$ neutrinos from the \emph{Fermi} bubbles over a four-year observation. In particular, this model cannot account for the observation of the PeV neutrino (Event 14) located within the \emph{Fermi} bubbles region, which is the least likely event to be produced through fluctuations of the astrophysical background.

\subsection{Leptonic-Hadronic Models of the \emph{Fermi} bubbles}\label{sec:lepHad}

In this scenario, we conversely assume that the bright GeV emission observed from the \emph{Fermi} bubbles is produced predominantly by leptonic interactions ($\ge 95\%$ below 100~GeV). Thus, the $\gamma$-ray spectrum from the hadronic portion of the \emph{Fermi} bubbles emission can be made arbitrarily hard, so long as it does not exceed HAWC limits. We note that harder $\gamma$-ray spectral indices will inevitably produce larger contributions to the IceCube neutrino flux, because IceCube observes messengers at higher energies than either the Fermi-LAT or HAWC. In the case of arbitrarily hard hadronic CR injection spectra, a large IceCube neutrino flux could be produced while remaining consistent with Fermi-LAT or HAWC upper limits. 

However, in this paper we adopt a reasonable lower limit on the CR proton energy spectrum of $\alpha$~=~2.0, motivated by the spectrum obtained via diffuse shock acceleration~\citep{1977ICRC...11..132A, Krimsky77, 1978MNRAS.182..147B, 1978ApJ...221L..29B}. We consider a pure power-law spectrum, and normalize the neutrino flux to  the neutrino data point.  Above $\sim$500~TeV, $\gamma$-ray spectrum is exponentially suppressed due to attenuation by the cosmic microwave background.
We note that an additional exponential cutoff in the injection spectrum around 7~PeV or less is required to respect the CASA-MIA $\gamma$-ray upper limit around 10~PeV.

We find that this hybrid leptonic-hadronic scenario is consistent with current HAWC upper limits. Moreover, the injected cosmic-ray spectrum can be slightly softened compared to the theoretical upper limit, before the observed $\gamma$-ray signal would saturate the HAWC upper bound. 

\subsection{Future Constraints from HAWC}
So far, our results employ less than one year of HAWC data. If future HAWC observations do not find emission consistent with the \emph{Fermi} bubbles, we expect these upper limits to fall as the square root of time. In Figure~\ref{fig:Nyr}, we show the maximum number of neutrinos that would be consistent with HAWC upper limits assuming $n$ additional years of HAWC data, and no detection of $\gamma$-ray emission from the \emph{Fermi} bubbles (red line). We calculate this upper limit using the maximum $\alpha$~=~2.0 spectrum and normalize the $\gamma$-ray flux to the center of the last bin in the HAWC data, which is located at 69.5~TeV. 
Additionally, we show the {\it total} number of IceCube events predicted in the \emph{Fermi} bubbles region, given $n$ more years of IceCube data\footnote{We note that IceCube has  recorded   more years of HESE data than have been publicly released, so the improvement in data can not be placed on a standard timeline for both HAWC and IceCube observations.}. 
In particular, we provide three models which bound our current knowledge of IceCube observations. First, we provide a prediction for the total neutrino flux from the \emph{Fermi} bubbles region, based on an extrapolation of the number of HESE events that overlap with the bubbles in the 4-year data (black dashed line). Second, we show a projection based on the best-fit flux of IceCube neutrinos from the \emph{Fermi} bubbles as calculated in this paper (see Sec.~\ref{sec:flux}). This translates to $N_{\rm FB}=1.03$ events over four years of data. Third, we show a projection based on an IceCube neutrino flux which saturates the 90\% confidence upper-limit on the \emph{Fermi} bubbles contribution to the IceCube neutrino flux, and which is predicted to produce $N_{\rm FB}=5.99$ events in four years of data. In this case, we note that the first four-years of data must include an anomalously low flux of neutrinos produced by the \emph{Fermi} bubbles.

We find that, at present, a $\gamma$-ray flux that saturates the HAWC upper limits in our hybrid leptonic-hadronic  model produces $N_{\nu,\rm 4yr}^{\rm max} = 10.5$ neutrinos in the \emph{Fermi} bubbles. This exceeds the 5 neutrinos that currently overlap with the \emph{Fermi} bubbles region, and indicates that HAWC can not currently rule out a \emph{Fermi} bubbles origin for the IceCube neutrinos in the hybrid leptonic-hadronic model.
However, the number of allowable neutrino events falls approximately as the square-root of time, as additional HAWC constraints more stringently rule out a sizeable IceCube neutrino flux. Eventually, we note that this upper-limit becomes flat, as the additional flux sensitivity of HAWC is offset by the increase in the IceCube acceptance over time.

Figure~\ref{fig:Nyr} shows that if HAWC does not detect $\gamma$-ray emission from the \emph{Fermi} bubbles within seven additional years of operation, approximately 50\% of the neutrino events in the bubbles region can be excluded as having a bubbles origin. While we note that this constraint appears weak, it is based solely on the synergy between very-high-energy $\gamma$-rays and high-energy neutrinos, and thus has independent systematics from our models of the IceCube background. If the true neutrino flux from the \emph{Fermi} bubbles is instead represented by the best fit event rate of $N_{\rm FB}=1.03$ events per four years of HESE data, HAWC will be unable to substantially constrain this model within 15 years of data. However, if the true neutrino event rate from the Fermi bubbles saturates the current IceCube upper limit, HAWC should begin to observe, or constrain a bubbles origin of this emission within $\sim$5~years. This indicates that HAWC observations are capable of testing currently viable models for the \emph{Fermi} bubbles, even for extremely difficult hybrid leptonic-hadronic models. In particular, we note that throughout this section, we have assumed an E$^{-2}$ spectrum for both the hadronic $\gamma$-ray and neutrino flux in the hybrid leptonic-hadronic model. Softer spectral indices will produce more stringest limits from existing HAWC data.


\section{Discussion}\label{sec:discussion}

We show that in four-years of IceCube data, eight HESE shower events overlap with the \emph{Fermi} bubbles region. Five of those events are centered within the \emph{Fermi} bubbles. The flux and spectrum of these events are consistent with expectations from a combination of atmospheric and astrophysical background neutrinos (p-value 0.22), allowing us to set strong upper limits on the maximum neutrino flux produced by the \emph{Fermi} bubbles. These observations produce  tension with models where the \emph{Fermi} bubbles are produced by pure hadronic processes, indicating that the CR proton spectrum in these models must be exponentially suppressed above energies of $\sim$100~TeV.  

While IceCube observations are likely to provide increasingly stringent constraints on hadronic models of the \emph{Fermi} bubbles, our results indicate that future HAWC observations may be able to place even stronger constraints on the maximum neutrino flux from the bubbles. Already, the combination of Fermi-LAT and HAWC $\gamma$-ray observatinos rule out models where a significant IceCube neutrino flux is produced by the \emph{Fermi} bubbles in a pure hadronic model. In the case of a hybrid leptonic-hadronic model, the constraining power of HAWC observations relies on the assumed spectrum of the hadronic emission component. However, even for CR spectral indices as hard as $\alpha$~$\sim$~2.0, null-observations with a few more years of HAWC data will strongly constrain models where the IceCube neutrino flux stems from true \emph{Fermi} bubbles emission.

The total cosmic-ray energy in a hybrid leptonic-hadronic scenario can be estimated as follows. Fitting to the IceCube data point suggests a gamma-ray flux  \mbox{$J_\gamma \lesssim 2\times10^{-8}\,\rm GeV\,cm^{-2}\,s^{-1}\,sr^{-1}$,}
corresponding to an integrated gamma-ray flux $2.7\times10^{-7} \,\rm GeV \,cm^{-2}\, s^{-1}\, sr^{-1}$ before attenuation, assuming an $E^{-2}$ spectrum and $E_{p,\rm cut} = 7$~PeV as discussed in Sec.~\ref{sec:lepHad}.  Then the $\gamma$-ray luminosity is \mbox{$L_\gamma= 3.9\times10^{36}\,\left(R/9.4{\,\rm kpc}\right)^2\,\rm erg\,s^{-1}$,} where $R$ is the distance to the bubbles. This implies a total CR proton energy $W_p \lesssim 2\,t_{\rm pp}\,L_\gamma=4.4\times10^{53}\,\left(R/9.4{\,\rm kpc}\right)^2\,\left(n_H/0.01\,\rm cm^{-3}\right)^{-1}\,\rm erg$. The energy contained in the electron population needed to produce the bulk of the observed GeV $\gamma$-ray flux is $W_e=10^{52}\,\rm erg$ \citep{FB2014}. This leads to a CR electron-to-proton ratio  $W_e/W_p\gtrsim 0.02$, which is consistent with the value of $10^{-3}-10^{-2}$ expected from diffusive shock acceleration \citep{2015PhRvL.114h5003P}. Future HAWC observations may be able to resolve this hadronic component based on its TeV emission.

Finally, we note that future instruments, such as the Large High Altitude Air Shower Observatory (LHAASO)~\citep{doi:10.1142/S2010194512005867, 2016arXiv160207600D}, the Hundred Square km Cosmic ORigin Explorer (HiScore)~\citep{2012NIMPA.692..246T}, and the Cherenkov Telescope Array (CTA)~\citep{2011ExA....32..193A} will soon provide additional high-sensitivity observations of the \emph{Fermi} bubbles, allowing the combined $\gamma$-ray constraints from these models to strongly constrain the spectrum and intensity of the $\gamma$-ray and neutrino emission from the bubbles.

\section*{Acknowledgements}
We thank E. Blaufuss, D. Fiorino, C. Kopper, C. Rivi\`ere, D. Malyshev,  and A. Smith    for helpful comments.
K. F. acknowledges the support of a Joint Space-Science Institute prize postdoctoral fellowship.
TL acknowledges support from NSF Grant PHY-1404311. KM is supported by NSF Grant No. PHY-1620777.

\bibliography{bubble}
\end{document}